\documentstyle[aps]{revtex}
\newcommand{\subs}[1]{\mbox{\scriptsize \it #1}}
\draft 
\begin{document}
\title{Genetic Algorithm Dynamics on a Rugged Landscape}   
\author{Stefan Bornholdt}
\address{Institut f\"ur Theoretische Physik, Universit\"at Kiel \\
Leibnizstrasse 15, D-24098 Kiel, Germany \\
bornholdt@theo-physik.uni-kiel.de\\
\date{Received 8 April 1997; revised manuscript received 5 December 1997}   
}    
\maketitle

\begin{abstract}  
The genetic algorithm is an optimization procedure motivated by biological evolution  
and is successfully applied to optimization problems in different areas.   
A statistical mechanics model for its dynamics is proposed based 
on the parent-child fitness correlation of the genetic operators, 
making it applicable to general fitness landscapes. 
It is compared to a recent model based on a maximum entropy ansatz. 
Finally it is applied to modeling the dynamics of a genetic algorithm on the 
rugged fitness landscape of the NK model. 
\medskip \\ 
PACS numbers: 05.50.+q, 87.10.+e, 07.05.Mh, 02.60.Pn 
\end{abstract}
\pacs{Published in Phys.\ Rev.\ E {\bf 57}, 3853 (1998).}   

\section{Introduction}  
The idea of utilizing biological evolution \cite{darwin} as a metaphor for an optimization algorithm 
is not recent \cite{ga1} - practical implementations, however, had to wait until 
computers of adequate speed were available \cite{ga2}. Finally, recent advances in 
this technology jump-started a surge of evolutionary approaches to optimization problems 
in real world applications \cite{ga_appl}. Applications to theoretical problems include the 
ground state search in some spin glasses \cite{sg}. Despite this renewed activity, the dynamics of 
these algorithms is not nearly as well understood as of other common optimization 
techniques, e.g., simulated annealing \cite{sa}. 
As one widely used representative of the biologically motivated optimization algorithms, the 
``genetic algorithm'' bases its search on a set of search points ( a ``population'' in the biological picture). 
A dynamical rule constructs a new population of search points from it, in a way that on average its energy 
decreases (or ``fitness'' increases) with respect to the given optimization function. 
The dynamical operators include selection and reproduction of the fittest members, as well as
mutation and recombination of members to generate new search points. 
One reason for the difficulty in modeling this algorithm 
is the non-gradient nature of the search space exploration allowing for 
non-local moves which complicates the treatment of its dynamics 
as a Markov chain \cite{markov}.       

A common approach to modeling physical systems with a large number of 
degrees of freedom is to find a few macroscopic variables that 
describe the average behavior of a system (e.g., temperature for 
a gas of a large number of atoms). In equilibrium systems, there are canonical 
procedures to describe these variables. In systems far from equilibrium,
as are genetic algorithms, one can sometimes identify distributions 
that tend to become stationary under the dynamics. 
Recently, a theoretical approach to the dynamics of the evolving, 
finite size population of a genetic algorithm has been proposed that 
uses the fitness distribution of the population as the characteristic evolving
quantity \cite{pbs1}. While   
this is successfully applied to selected, simple optimization problems \cite{pbs2}, the method becomes 
difficult for problems that are more complex, since it depends on an intricate maximum likelihood 
estimation. 
This is used to describe the dynamics of the genetic operators 
in terms of a fitness distribution, where structural 
effects have to be averaged over and re-expressed in terms of fitnesses. 
In this study we will go back one step and look at the lowest order observables 
that determine the dynamics. 
In particular, 
we construct a simplified model based on the observation that genetic 
algorithm performance often correlates strongly with the parent-child fitness correlation. 
While using the selection scheme of \cite{pbs1},
the correlation is used to construct a dynamical model  
which is applied to a simple additive fitness function as well as the spin glass motivated NK model. 
This will show how well the concept of the correlation determining the dynamics 
actually holds in modeling the dynamics of a genetic algorithm, and where its limits are.  

The motivation for our model stems from the observation that, although  
the time evolution of a genetic algorithm is difficult to understand, 
practitioners often have a quite distinct intuition about when a genetic algorithm will work well. 
One common statement in this respect is that the algorithm performs best when the fitnesses of 
parents and their children are strongly correlated.
It was found empirically 
that the performance of a genetic algorithm follows the fitness correlation 
of the genetic operators as well as the correlation length of the optimization landscape  
\cite{manderick}.    
In fact, designing suitable genetic representation schemes often means increasing the
parent-child fitness correlation. Finally, in biological systems the correlation 
between parents and children is usually very large.   
What does this intuition mean in the light of existing models for genetic algorithm dynamics? 
To what extent are correlation measures able to predict the algorithm dynamics? 
These are exactly the questions that we will study in the following. 

We will now first describe the simple version of a genetic algorithm, then review 
the correlation model and proceed with modeling the genetic operations on the  
basis of fitness correlations, first for mutation algorithms and finally for full genetic 
algorithms, including recombination.
The study concludes with a comparison of the model to numerical studies of the 
corresponding algorithms.  

\section{A Dynamical Model of the Basic Genetic Algorithm}  
A genetic algorithm performs population based optimization in a discrete search space. 
For our purposes we define a simple version of this algorithm. 
Let the function which is to be optimized be a real valued function $f(S)$   
on a binary search space representation $S \in \{\pm1\}^N$ of dimension $N$. 
Its value is the ``fitness'' of the test point $S$ which is to be maximized 
(alternatively one could view $-f(S)$ as an energy to be minimized).  
In the biological picture, $S$ is the analogue to the genome.
The algorithm starts from a random ``population'' of search points 
$S_\alpha$ with $\alpha =1,\dots,P$ forming a population of $P$ strings with fitnesses 
$f_\alpha = f(S_\alpha)$.  
Subsequently, new search points are tested by means of three operations, 
called selection, mutation, and recombination. 
In the selection step, a new population is created:  
members are selected according to probabilities defined on the basis of their fitness values.   
Those with a higher fitness are more likely to ``survive'' than those with smaller fitness
and the new population most likely has a higher average fitness than the old one. 
New search points are then created 
by flipping single spins with a small fixed probability $\gamma$ in all of the 
population, called the mutation step. Then pairs of strings are allowed to exchange a subset 
of their sites in a recombination step, analogous to the biological process of crossing over
genomes. This procedure is iterated resulting in an evolution towards higher fitness values.     
It can be used to solve optimization problems where the problem is defined by 
means of a scalar fitness function $f(S)$. The choice of the structure of the search space and 
the encoding of $f(S)$ often determine the convergence properties and performance of the 
algorithm and is one major motivation for the modeling of genetic algorithm dynamics. 

In the following, the dynamics of the genetic operators will be studied in more detail. 
We will do this 
for two toy models, an additive fitness function on the one hand, and a more 
rugged function on the other.  
The first problem is the simplest additive function, a random field paramagnet 
\begin{eqnarray}
f_\alpha = \sum\limits_{i=1}^N J_i\; S_i^\alpha + \kappa_1^0  
\end{eqnarray}
with random couplings $J_i$ taken from a Gaussian distribution 
with mean $0$ and variance $1$. The $N$ sites $S_i^\alpha$ with 
$i=1,\dots,N$ and $S_i^\alpha = \pm 1$ form the genetic string 
of the member $\alpha$ of the population.
The second function will be the NK model fitness function\cite{nkmodel}  
\begin{eqnarray}
f_\alpha = \sum\limits_{i=1}^N E_i(S_i^\alpha;S_{i_1}^\alpha,
           \dots , S_{i_K}^\alpha)   
\label{nkfunc}  
\end{eqnarray}
with $2^{K+1}$ random energy values $E_i(S^\alpha)$ drawn from a uniform 
distribution over the interval $[0,1]$ and a randomly chosen permutation of  
sites $i_1$ to $i_K$, both for each $i$. 
Originally, this function has been formulated for the study of evolution on tunably 
rugged fitness landscapes with application to the evolution of the immune response 
\cite{immune}.  
For these two functions, let us derive the dynamics under mutation and recombination. 

The dynamics of the model is described in terms of the fitness
distribution $\rho(f)$ of the population which is expressed as
an expansion in cumulants as proposed in \cite{pbs1}. 
The cumulants $\kappa_n$ of $\rho(f)$ are defined through
\begin{eqnarray}
\kappa_n =
\frac{\partial^n}{\partial \beta^n} 
\int\limits_{-\infty}^\infty \exp(\beta f) \; \rho(f) \; df
\Bigg|_{\beta=0}, 
\end{eqnarray}
representing the mean, variance, skew ($\kappa_3/\kappa_2^{3/2}$),  
curtosis ($\kappa_4/\kappa_2^2$), and higher moments of the fitness distribution.
To give an intuitive picture, the first two cumulants roughly capture
the infinite population size limit of the model.
The higher cumulants, skew and curtosis, are important to describe the
dynamics of a finite population where, e.g., selection causes the fitness
distribution to quickly become skewed and thus deviate from a Gaussian.
An evolving population can, at each time step, be
approximated by a set of these variables. Its dynamics can then be
viewed in terms of the evolution of the cumulants.
In the following, the dynamics of an evolving population will be
modeled using a truncated expansion in the first four cumulants.
The different operations of a genetic algorithm, selection, mutation,
and recombination, interact in different ways with this representation.

First consider selection. We will follow the formalism of \cite{pbs2} for modeling the 
selection operation on a fitness distribution in terms of cumulants. 
Boltzmann selection is considered where  
a member with fitness $f_\alpha$ is chosen from the population
with the probability
\begin{eqnarray}
p_\alpha = \frac{e^{\beta f_\alpha}}{Z}, \; \; \;
Z = \sum\limits_{\alpha=1}^Pe^{\beta f_\alpha}
\end{eqnarray}
where $\beta$ parametrizes selection strength.
After selection, the cumulants are given by
\begin{eqnarray}
\kappa_n^s = \frac{\partial^n}{\partial\beta^n} < \ln Z >_\rho
\end{eqnarray}
where the average is taken over all possible populations with individual
fitnesses satisfying the given $\rho(f)$. This expression can be
solved similarly to the random energy model \cite{rem}, as shown in \cite{pbs2}.
One obtains the cumulants after selection as functions
of the cumulants before selection, either by means of numerical integrals or,
in the limit of small selection (small $\beta$), as an expansion. 
The cumulants after selection have been derived under the assumption
that the new population is drawn from a continuous
fitness distribution. Only the dominant finite-population effect is kept
which originates from the stochastic sampling of the new population in the
selection step.

While selection is solely determined by the fitness distribution
of the population, 
the other two operators act on the representation $S$ instead. 
The average effect of the mutation operator for the random field paramagnet
has been worked out in \cite{pbs2} by averaging over all possible mutation 
events yielding to lowest order  
\begin{eqnarray}
\kappa_1^i \equiv \left< f_\alpha^m \right>_{\subs mut} = m \; f_\alpha + (1 - m) \; \kappa_1^0 
\label{avfm}  
\end{eqnarray}
with  
\begin{eqnarray}
m  = 1-2\gamma. 
\end{eqnarray}
Similarly, we derive for the NK-model the fitness of a mutated string 
by writing down (\ref{nkfunc}) with each site 
$S_i^\alpha$ multiplied by a random $\sigma_i^\alpha = \pm 1$. 
The energy of a single $E_i$ changes 
to another (randomly chosen) value if at least one of the sites is changed and
remains unchanged otherwise. The average fitness of a string after 
mutation is then obtained in an annealed approximation as (\ref{avfm}) with 
\begin{eqnarray}
m = (1-\gamma)^{K+1}.   
\end{eqnarray}
For both functions we can write the mean fitness $\kappa_1^i$ of the potential children 
of a parent with fitness $f_\alpha$ as (\ref{avfm}) with some function dependent constant $m$. 
In general terms, $m$ is the fitness correlation of a genetic operator (here: mutation) 
with respect to a specific fitness landscape (here: $f$). 
The above observation motivates to use $m$ as a measure of the lowest order 
genetic algorithm dynamics on general landscapes. 
Defined in terms of an average 
over the population and possible mutation events, $m$ can also be expressed as    
\begin{eqnarray}
m = \frac{
        \left< f_\alpha f_\alpha^m \right>_{\alpha, mut}  
        - \left< f_\alpha \right>_\alpha \left< f_\alpha^m \right>_{\alpha, mut}  
    } {
        \left< f_\alpha^2 \right>_{\alpha} - \left< f_\alpha \right>_\alpha^2  
    }.        
\end{eqnarray}
It can thus be measured from a given fitness function. Here, it parametrizes 
the average fitness of a member after mutation (given the fitness before) and will 
be used below to give a lowest order approximation of the population dynamics. 

Let us check how the next order relates to this picture. 
The fitness variance of mutated members $f^m_\alpha$ 
derived from a single parent with fitness $f_\alpha$
has been calculated for the random field paramagnet in \cite{pbs2} as 
\begin{eqnarray}
\kappa_2^i &=& 
\left< (f^m_\alpha)^2 \right>_{\subs mut} - \left< f^m_\alpha \right>_{\subs mut}^2
\nonumber \\   
&=&   
(1-m^2) \; \sum\limits_{i=1}^N \; J_i^2.   
\end{eqnarray}
For comparison we obtain for the NK model by a similar calculation 
\begin{eqnarray}
\kappa_2^i = 
(1-m^2) \; \kappa_2^0
+ 
m \; (1-m) \; \left[ \sum\limits_{i=1}^N E_i^2(S^\alpha) 
- \left( \kappa_2^0 + \frac{(\kappa_1^0)^2}{N} \right)   
- \frac{2}{N} \; \kappa_1^0 \; \left( f_\alpha -\kappa_1^0 \right) 
\right] 
.   
\end{eqnarray}
Other that the first order terms, these expressions are not exactly of an equal type.   
However, looking for a lowest order rule to approximate the variance
after mutation, let us consider the average over the class of allowed functions.
Averaging over $J$ resp.\ $E$ for the two problems, 
properties of a particular realization drop out and one obtains 
\begin{eqnarray}
\left< \kappa_2^i \right> = (1-m^2) \; \kappa_2^0.    
\label{k2approx}  
\end{eqnarray}
This is again a function of the fitness correlation $m$, motivating 
a model of the genetic algorithm dynamics. 
This corresponds to modeling the distribution 
of fitness values after mutation from a parent of fitness $f_\alpha$ by the ansatz  
\begin{equation}
\rho(f^m | f) = \frac{1}{\sqrt{2\pi\kappa_2^i}} \; 
\exp\left(-\frac{(f-\kappa_1^i)^2}{2\kappa_2^i}\right)  
\label{mutcond}  
\end{equation}
with  
\begin{eqnarray}
\kappa_1^i &=& m \; f_\alpha +  (1-m) \; \kappa_1^0
\nonumber \\  
\kappa_2^i &=& (1-m^2) \; \kappa_2^0  
\label{mutmod}  
\end{eqnarray}
where $\kappa_1^0$ and $\kappa_2^0$ are the cumulants of the initial, random  
distribution. 
It reflects the empirical observations about genetic algorithm performance on
correlated landscapes \cite{manderick}.  
With this ansatz, the fitness distribution 
after mutation is predicted as 
\begin{eqnarray}
\kappa_1^m &=& m \; \kappa_1  + (1 - m) \; \kappa_1^0    
\nonumber \\
\kappa_2^m &=& m^2 \; \kappa_2  + (1 - m^2) \; \kappa_2^0    
\nonumber \\
\kappa_3^m &=& m^3 \; \kappa_3  
\nonumber \\
\kappa_4^m &=& m^4 \; \kappa_4.  
\label{modmut}
\end{eqnarray}
This is a lowest order model for mutation dynamics based on a given 
parent-child fitness correlation $m$. 
To compare this prediction with a direct calculation from the fitness functions, 
the distribution of the population after mutation 
is obtained by an additional average over the parents in all possible populations. 
Neglecting finite-population effects in the mutation step which are much smaller 
than those in selection, the first cumulant of 
the distribution of the population after mutation is then  
\begin{eqnarray}
\kappa_1^m = m \; \kappa_1 + (1 - m) \; \kappa_1^0    
\end{eqnarray}
for the random field paramagnet as derived in \cite{pbs2}.   
We obtain the same expression for the NK-model. 
The second order of the random field paramagnet has been derived as  
\begin{eqnarray}
\kappa_2^m = m^2 \; \kappa_2 + \; (1 - m^2) \; \sum\limits_{i=1}^N \; J_i^2.      
\end{eqnarray}
In comparison we obtain for the the NK model   
\begin{eqnarray}
\kappa_2^m = 
m^2 \; \kappa_2 
&+& 
(1 - m^2) \; \kappa_2^0 
\nonumber \\   
&+& 
m \; (1-m) \; 
\left[ \sum\limits_{i=1}^N \left< E_i^2(S^\alpha) \right>_\alpha 
- \left( \kappa_2^0 + \frac{(\kappa_1^0)^2}{N} \right) - \frac{2}{N} \; \kappa_1^0 \; 
\left( \kappa_1 - \kappa_1^0 \right)
\right].  
\label{k2mnk}  
\end{eqnarray}
The full second-order expression cannot be derived from a knowledge of  
$m$ alone, due to fluctuations in the third term of (\ref{k2mnk}). 

In general, one finds that the cumulants after mutation do not 
always depend on the pure cumulants before selection.   
The dynamics also depends on properties of the genetic coding since 
it directly acts on the underlying representation. 
If the fitness distribution is all one knows about a population, one
has to make additional assumptions when modeling the dynamics in order 
to describe the underlying dynamics of the genetic variables correctly.
The model developed in \cite{pbs2} utilizes a maximum entropy estimation
for this purpose. Since this is a complicated method for general hard 
optimization problems, we here use a different approach by concentrating
on a lowest order dynamical model of the basis of the correlation $m$. 
This method is more accessible for complicated fitness landscapes. 
In the example of the rugged NK landscape we find that the first two 
cumulants after mutation are well reproduced in terms of $m$ with the fluctuations
in (\ref{k2mnk}) being small.   
Therefore, we approximate the second cumulants of both above examples 
by an expression as done in (\ref{modmut}).    
This set of equations derives 
the mutation cumulants from a model for a ``microscopic'' 
mutation event (\ref{mutcond}) on the basis of the fitness correlation of mutation applied 
to the landscape. 
Applying mutation to a member $\alpha$ of the population with initial 
fitness $f_\alpha$, the resulting member will in general have a different 
fitness value $f_\alpha^m$. The case of maximum correlation 
$f_\alpha^m = f_\alpha$ occurs where mutation does not affect the fitness of 
the child at all. 
In the other extreme of violent mutation, it involves a random change in fitness,   
leaving traces of the fitness distribution of a {\em random} population $\rho^0(f_\alpha^m)$.   
We parametrized the possible correlations in the range between these two 
extremes by approximating the fitness distribution of the child as (\ref{mutmod}). 
In this way, the degree of correlation between parent and child under mutation 
is parametrized by $m$.
The choice of the two distributions is natural since, in general, $\rho^0$ 
is the fixed point distribution of the mutation operator. 
On this basis, (\ref{modmut}) defines a closed expression for the fitness 
distribution of the population after mutation as a function of 
the distribution prior to mutation. It will serve as an iteration 
step in describing the complete dynamics.  

A similar approach as for mutation can be adopted for recombining genetic strings. 
In the simple genetic algorithm modeled here, the recombination operation is
defined symmetrically in the two children produced.  
\begin{enumerate}   
\item Group the individuals of a population into pairs of two. 
\item Recombine each pair, i.e., at each Bit position, 
      swap the adjacent spins with a given probability $a$. 
\item For each pair, replace the parents by the two children thus produced.  
\end{enumerate}   

For the random field paramagnet model, according to \cite{pbs2} the fitness 
of one child produced by recombination averaged over all possible recombination events is then  
\begin{eqnarray}
\left< f^c_{\alpha\beta} \right>_{\subs cross} = a \; f_\beta + (1-a) \; f_\alpha.    
\end{eqnarray}  
Along similar lines, for the NK model, let us write the fitness of a child 
averaged over all possible recombination events as 
\begin{eqnarray}
\left< f_{\alpha\beta}^c \right>_{\subs cross} &=& 
\sum\limits_{i=1}^N\; \sum\limits_{n=0}^{K+1}\; 
{K+1 \choose n}\; a^n \; (1-a)^{K+1-n} \;  
\Bigg\{ \frac{1}{2} \; 
\left[ E_i(\vec S^\alpha) + E_i(\vec S^\beta) \right] \; q^{K+1} 
\nonumber \\   
&+& E_i(\vec S^\alpha) \; q^n\left(1-q^{K+1-n}\right) 
+ E_i(\vec S^\beta) \; q^{K+1-n}\left(1-q^n\right) 
\nonumber \\   
&+& \frac{1}{2} \; \left(1-q^n\right)\left(1-q^{K+1-n}\right)
\Bigg\}.     
\label{crossednk}  
\end{eqnarray}
Here, $n$ denotes the number of sites swapped between the
arguments of one corresponding energy term $E_i$ of the parents. 
The sum over $n$ is followed by the probability of exactly $n$ swapped sites 
in a set of $K+1$ sites relevant for each energy term. 
The last term is the annealed average of a child's energy term $E_i$, where $q$ is the 
average probability that two random sites $S^\alpha_i$ and $S^\beta_i$
are equal in the population. For both fitness functions, the post-recombination 
fitness can be cast into the unified expression   
\begin{eqnarray}
\left< f_{\alpha\beta}^c \right>_{\subs cross} = 
c_{\alpha\beta} \; f_\alpha 
+ c_{\beta\alpha} \; f_\beta 
+ (1 - c_{\alpha\beta} - c_{\beta\alpha}) \; \kappa_1^0 
\label{prf}  
\end{eqnarray}
with $c_{\alpha\beta} = a$, $c_{\beta\alpha} = 1-a$ for the random field paramagnet, and 
\begin{eqnarray}
c_{\alpha\beta} &=& \left[ 1 - a + a \; q \right]^{K+1} - {\textstyle \frac{1}{2}} \; q^{K+1}  
\nonumber \\ 
c_{\beta\alpha} &=& \left[ a + (1-a) \; q \right]^{K+1} - {\textstyle \frac{1}{2}} \; q^{K+1}  
\label{cnk}  
\end{eqnarray}
with $\kappa_1^0 = N/2$ for the NK model. 
The parameters $c_{\alpha\beta}$ and $c_{\beta\alpha}$ correspond to the average 
fitness correlations of a child $f_{\alpha\beta}^c$ with either one of its 
parents, $f_\alpha$ or $f_\beta$, after recombination: 
\begin{eqnarray}
c_{\alpha\beta} &=& \frac{ 
\big< f_\alpha f_{\alpha\beta}^c \big>_{\alpha \not= \beta, cross} - \big< f_\alpha \big>_{\alpha}
  \big< f_{\alpha\beta}^c \big>_{\alpha \not= \beta, cross}
}{\left( 1 - \frac{1}{P-1} \right) \; \kappa_2}  
\nonumber \\
c_{\beta\alpha} &=& \frac{ 
\big< f_\beta f_{\alpha\beta}^c \big>_{\alpha \not= \beta, cross}
- \big< f_\beta \big>_{\beta}
  \big< f_{\alpha\beta}^c \big>_{\alpha \not= \beta, cross}
}{\left( 1 - \frac{1}{P-1} \right) \; \kappa_2}.  
\label{ccx}
\end{eqnarray}
Here, the averaging is done over all members in the population and, after 
recombination, over all possible pairings and all possible recombination events. 
From the post-recombination fitness (\ref{prf}) we can again derive the average 
fitness of the population after recombination by averaging over all potential parents,   
yielding 
\begin{eqnarray}
\kappa_1^c = c \; \kappa_1  + (1 - c) \; \kappa_1^0      
\label{lo}  
\end{eqnarray}
with $c = c_{\alpha\beta} + c_{\beta\alpha}$.   
For the first order cumulants we find that both functions, the random paramagnet 
and the NK model, fall into the same model class. 
This will motivate us below to use the correlation $c$ for a lowest order dynamical
model of recombination. 
The higher moments can be derived in a similar fashion. 
We define the fitness variance of the population after recombination as 
\begin{eqnarray}
\kappa_2^c &=& \left< (f_{\alpha\beta}^c)^2 \right>_{\alpha \not= \beta, cross} 
              - \left< f_{\alpha\beta}^c \right>_{\alpha \not= \beta, cross}^2    
\end{eqnarray}
as the variance of an infinite size population of children 
derived from a finite parent population and averaged over all allowed recombination events 
and parental pairs. 
This has been calculated in \cite{pbs2} for the random field paramagnet as 
\begin{eqnarray}
\kappa_2^c &=& \kappa_2. 
\end{eqnarray}
In this case, recombination leaves mean and variance of fitness untouched.     
The third moment is defined as  
\begin{eqnarray}
\kappa_3^c &=& \left< f_{\alpha\beta}^{c3} \right>_{\alpha \not= \beta, cross} 
-3\kappa_1^c\kappa_2^c - \kappa_1^{c3}  
\end{eqnarray}
and, dropping spatial correlations, is given in \cite{pbs2} as 
\begin{eqnarray}
\kappa_3^c &=& 
\left[ a^3 + (1-a)^3 \right] \; \kappa_3 
- 6 a(1-a) \; \sum\limits_{i=1}^N \; J_i^3 \; \left\{ \big< S_i^\alpha  \big>_\alpha  
- \big< S_i^\alpha S_i^\beta S_i^\gamma \big>_{\alpha, \beta, \gamma} \right\}.   
\label{third}  
\end{eqnarray}
One now obtains also terms containing spin correlations. In the model of \cite{pbs2} 
these are estimated in a maximum entropy estimation, summing over search space 
regions corresponding to a given fitness distribution of the population. 
As this approach becomes again impractical for real optimization problems 
with hard or unknown fitness functions, we here study the simpler approach  
to describe recombination dynamics on the basis of the fitness correlation $c$. 
We will see below that this works well in cases where the fluctuations from 
spin correlations remain small as in the case of the random field paramagnet. 
For comparison let us also consider recombination of the more difficult landscape 
of the NK model.  
For the variance we obtain   
\begin{eqnarray}
\kappa_2^c &=& \left( c_{\alpha\beta} + c_{\beta\alpha} \right) \; \kappa_2 
+ \left( 1 - c_{\alpha\beta} - c_{\beta\alpha} \right) \; \kappa_2^0 
\nonumber \\  
&+& \left( c_{\alpha\beta} + c_{\beta\alpha} \right) \; \left( 1 - c_{\alpha\beta} - c_{\beta\alpha} \right) 
\; \left\{ \kappa_1^2 + \frac{1}{N} \; \left[ (\kappa_1^0)^2 - 2\kappa_1\kappa_1^0 \right] \right\}   
\nonumber \\  
&+& \sum\limits_{i=1}^N \; \sum\limits_{j \not= i} \;  
\left\{ \left( c^2_{\alpha\beta} + c^2_{\beta\alpha} - c_{\alpha\beta} - c_{\beta\alpha} \right)   
\; \big< E_i^\alpha E_j^\alpha \big>_\alpha   
+ 2 \; c_{\alpha\beta} \; c_{\beta\alpha}    
\; \big< E_i^\alpha E_j^\beta \big>_{\alpha \not= \beta}  \right\}.      
\end{eqnarray}
This is now a different situation than before: The third term contains large fluctuations from correlations  
between energy terms within strings and within the population. In this case there is no strong limit of 
vanishing spatial correlations, instead $\big< E_i^\alpha E_j^\alpha \big>_\alpha - \big< E_i^\alpha \big>_\alpha 
\big< E_j^\alpha \big>_\alpha \not= 0$ for $i\not= j$, such that $\big< E_i^\alpha E_j^\alpha \big>_\alpha \approx  
\; \big< E_i^\alpha E_j^\beta \big>_{\alpha \not= \beta}$ is only weakly fulfilled. 
This is due to each energy term $E_i$ being coupled to neighboring
spins in the string. When running a real genetic algorithm with $a=1/2$ (as often used) one can 
observe the fluctuations with magnitudes comparable to the leading terms and of either signs. The very last term contributes 
also in the limit of large correlations $c_{\alpha\beta}$ and $c_{\beta\alpha}$ where the preceding 
term is suppressed. Here, any simple approximation breaks down, as does our intuition about a 
correlation governing the evolution. 
In fact, one is lead to consider that it might not be 
a question of a working description, rather than 
the issue of whether recombination helps at all in this limit. 
It clearly is disruptive here  
resulting in low correlation, a limit never encountered in biological evolution.      
For the NK model we will therefore 
consider here the less disruptive case of asymmetric recombination, 
resulting in better genetic algorithm performance, 
let us choose $a=1/2N$. In this limit $c_{\alpha\beta} \gg  c_{\beta\alpha}$ and the variance can now 
be written in the simple form  
\begin{eqnarray}
\kappa_2^c &=& c \; \kappa_2 + ( 1 - c ) \; \kappa_2^0 + c \; ( 1 - c ) 
\; \left\{ {\textstyle \left( 1 - \frac{1}{N} \right)} \; \kappa_1^2 
- \sum\limits_{i=1}^N \; \sum\limits_{j \not= i} \; \left< E_i^\alpha E_j^\alpha \right>_\alpha   
+ \frac{1}{N} \; \left( \kappa_1 - \kappa_1^0 \right)^2 \right\}  
\end{eqnarray}
with $c=c_{\alpha\beta}+c_{\beta\alpha}$. The remaining correlation is now balanced with 
$\left( 1 - \frac{1}{N} \right) \; \kappa_1^2$ and the difference suppressed by $(1-c)$.   
In this case, the lowest order model is  
\begin{eqnarray}
\kappa_2^c &=& c \; \kappa_2 + ( 1 - c ) \; \kappa_2^0.   
\end{eqnarray}
Now having the next to lowest order behavior of the two functions at hand, 
with the identical lowest order term (\ref{lo}), let us again use this 
as a motivation for a dynamical model based on $c$,    
as done before in the case of mutation. 
For this purpose we approximate the distribution after recombination by the
conditional probability density 
\begin{equation}
\rho(f^c_{\alpha\beta} | f_\alpha, f_\beta) = 
\frac{1}{\sqrt{2\pi\kappa_2^{ci}}} \; 
exp\left(-\frac{(f-\kappa_1^{ci})^2}{2\kappa_2^{ci}}\right)  
\label{crosscond}  
\end{equation}
with suitable moments $\kappa_1^{ci}$ and $\kappa_2^{ci}$. The mean fitness of the children 
of a given pair of parents is  
\begin{eqnarray}
\kappa_1^{ci} = 
c_{\alpha\beta} \; f_\alpha + c_{\beta\alpha} \; f_\beta + (1-c) \; \kappa_1^0 
\end{eqnarray}
as motivated by (\ref{prf}) and matches the fitness correlation picture as for mutation. 
Dealing with two parents with in general different fitness values, the recombination
event introduces also a variance. Let us first consider the random field paramagnet. 
Here one finds 
\begin{eqnarray}
\big< f_{\alpha\beta}^{c2} \big>_{cross} - \big< f_{\alpha\beta}^c \big>_{cross}^2    
= \sum\limits_{i=1}^N \; J_i^2 \; (1-S_i^\alpha S_i^\beta).  
\end{eqnarray}
Therefore, and since spatial correlations vanish here, the variance of the distribution 
of potential children can be modeled by 
\begin{eqnarray}
\kappa_2^{ci} = 
(c_{\alpha\beta} \; f_\alpha - c_{\beta\alpha} \; f_\beta)^2 + (1-c^2) \; \kappa_2^0.   
\end{eqnarray}
With this assumption, and with $c_{\alpha\beta} =  c_{\beta\alpha}$ for the random field paramagnet,  
the fitness distribution after recombination $\rho(f_{\alpha\beta}^c)$ is predicted as 
\begin{eqnarray}
\kappa_1^c &=& c \; \kappa_1  + (1 - c) \; \kappa_1^0    
\nonumber \\
\kappa_2^c &=& c^2 \; \kappa_2  + (1 - c^2) \; \kappa_2^0    
\nonumber \\
\kappa_3^c &=& c^3 \; \kappa_3.    
\end{eqnarray}
The mean and variance of the population after recombination are therefore correctly 
predicted for the random field paramagnet. The higher orders are off by a constant factor, they cannot 
be matched exactly within the second order correlation model, which would require
higher moments $\kappa_3^{ci}$ and $\kappa_4^{ci}$. 
The fixed point 
distribution for recombination (which does not equal that of a random population as for mutation) 
is small enough here to allow for neglecting the fluctuations in (\ref{third}) and higher 
moments. We will use this set of cumulants for the numerical model below.   

What happens in the case of the NK model? Here, the exchange of spins between the 
genomes has rather the effect of mutations than the sharing of knowledge 
between them. Using the same arguments as in the mutation model (\ref{mutmod}) we choose:
\begin{eqnarray}
\kappa_1^{ci} &=& c_{\alpha\beta} \; f_\alpha + c_{\beta\alpha} \; f_\beta + (1-c) \; \kappa_1^0 
\nonumber \\ 
\kappa_2^{ci} &=& (1-c^2) \; \kappa_2^0.    
\end{eqnarray}
The model then predicts in the asymmetric case of $c \approx  c_{\alpha\beta}$   
a post-recombination population distributed according to (\ref{modmut}), 
while the direct calculation suggests   
\begin{eqnarray}
\kappa_n^c &=& c \; \kappa_n  + (1 - c) \; \kappa_n^0.      
\end{eqnarray}

The correlation model, therefore, correctly predicts the mean fitness of the 
population after recombination, however, deviates in the higher orders. 
It still is numerically close to the direct calculation model which will be 
simulated up to $n=4$ below. 
An alternative microscopic model of recombination that predicts all leading orders to be 
linear in $c$ was proposed in \cite{sb}, however, does not improve the case under consideration here. 
The sets of cumulants from selection $\kappa_n^s$, mutation $\kappa_n^m$, and recombination $\kappa_n^c$, 
now define the iteration step of one ``generation'' of the dynamical model, 
with the operations being applied in this order. 

\section{Numerical comparison of the model to a genetic algorithm}  
In numerical simulations the correlation model is now be compared to the evolution of a real 
genetic algorithm.  The fitness distribution 
of the initial, random population for the random field paramagnet function is given by 
\begin{eqnarray}
\kappa_1^0 &=& \left< f_\alpha \right> _{\alpha, S, J} = 0
\nonumber \\
\kappa_2^0 &=& \left< \left< f_\alpha^2 \right>_\alpha 
            - \left< f_\alpha \right>^2_\alpha 
            \right>_{S, J} 
            = {\textstyle \left( 1-\frac{1}{P} \right)} \; N   
\nonumber \\
\kappa_3^0 &=& 0   
\nonumber \\
\kappa_4^0 &=& 
{\textstyle    
-6 \; N^2 \; \frac{1}{P} \; \left( 1 - \frac{1}{P} \right) \; \left( 1 - \frac{2}{P} \right) 
-6 \; N \; \left( 1 - \frac{1}{P} \right) \; \left( 1 - \frac{4}{P} + \frac{2}{P^2} \right) 
}.   
\end{eqnarray}
Since the average dynamics of a whole class of functions is 
considered here, the last value differs from the pure ensemble 
average $\left< \dots \right>_S$ through the additional average over all 
possible functions $\left< \dots \right>_J$.  
The corresponding results for the NK model are   
\begin{eqnarray}
\kappa_1^0 &=& \left< f_\alpha \right> _{\alpha, S, E} = \frac{N}{2}    
\nonumber \\
\kappa_2^0 &=& 
{\textstyle    
\left< \left< f_\alpha^2 \right>_\alpha - \left< f_\alpha \right>^2_\alpha \right>_{S, E} 
= \left( 1-\frac{1}{P} \right) \; \left( 1-\frac{1}{2^{K+1}} \right) \; \frac{N}{12}  
}   
\end{eqnarray}
and, omitting terms of orders $2^{-K}$,   
\begin{eqnarray}
\kappa_3^0 &=& 0   
\nonumber \\
\kappa_4^0 &=& 
{\textstyle    
- \left( 1 - \frac{1}{P} \right) \; \left( 1 - \frac{6}{P} + \frac{6}{P^2} \right) \; \frac{N}{120}  
- \frac{1}{P} \; \left( 1 - \frac{1}{P} \right) \; \frac{N^2}{24}. 
}   
\end{eqnarray}
The simulation results are averaged over 10000 runs of a 
genetic algorithm with population size $P=50$ and selection strength $\beta_s=0.01$  
(with a newly chosen random fitness function for each run). The size of the genetic 
string is $N=128$ sites and the mutation probability for each site is $\gamma = 1/2N$. 
In Fig.\ \ref{rp1}, the iterated cumulant expansion is compared to the dynamics of a 
genetic algorithm for the random field paramagnet with selection and mutation. 
The solid curves show mean and variance of the genetic algorithm fitness distribution
and are well described by the theoretical approach shown by the dashed curves. 
The theoretical model is based on the constant correlation value $m$. 
Although the correlation among genotypes in the population considerably changes over time, 
the fitness correlation $m$ remains in fact constant when measured in the population over time. 
Here, the correlation $m$ appears to contain the basic information about the dynamics. 
In Fig.\ \ref{nk1}, the evolution of the NK model fitness distribution 
for selection and mutation is shown for a model with $P=50$, $N=128$, and $K=8$.    
Again, the solid curves show the mean and variance of the measured 
genetic algorithm fitness distribution. 
The correlation $m$ is taken as derived above for the NK function.  
All other parameters are chosen as in the previous case.  
For the plot, $\kappa_1$ is depicted as $\kappa_1 - \kappa_1^0$.
As the figure shows, the evolution of the genetic algorithm is predicted correctly also on the 
rugged fitness landscape of the NK model.
The model yields a satisfactory prediction, especially when comparing  
the very simple correlation 
model to the maximum entropy model of \cite{pbs2}.  

Adding the recombination step to the simulation of the random field paramagnet, 
the modeling based on parent-child fitness correlations is shown in Fig.\ \ref{rp2}.  
Crossover has been defined to be ``uniform'', where each site is swapped 
with probability $a=0.5$ between the parents and both resulting children 
are taken. Here, the model uses the theoretical value of $c = 1$ while the 
remaining parameters of the simulations and the model are chosen as above.  
In the real genetic algorithm applied here to the random field paramagnet problem, recombination improves 
the performance as compared to Fig.\ \ref{rp1}. This is correctly predicted by the 
correlation model. 

While for the random field paramagnet we saw that $c$ remains constant over the 
course of evolution, for the NK model $c$ depends on the 
probability $q$ of two equal spins meeting in a recombination event. 
This is a quantity that cannot be expressed in terms of the fitness distribution 
alone. For the purpose of the numerical comparison we will estimate this  
probability from the average pair correlation in the population 
$\left< q_{\alpha\beta} \right>_{\alpha \not= \beta}$ 
with
\begin{equation}
q_{\alpha\beta} = \frac{1}{N}\sum\limits_{i=1}^N \; S_i^\alpha S_i^\beta 
\end{equation}
such that   
\begin{equation}
q = \frac{1+q_{\alpha\beta}}{2}.
\end{equation}
The average pair correlation is measured from the genetic algorithm runs 
and used to correct for the running $q$ in the numerical model. 
On the other hand, a closed model could easily be obtained by including $q_{\alpha\beta}$
as a dynamical variable into the models as proposed in \cite{magnus}.  
The result is shown in Fig.\ \ref{nk2} where the evolution of the NK-model fitness 
distribution for selection and recombination is shown using a selection strength $\beta_s=0.01$ 
and asymmetric crossover with $a=1/2N$ (again, $\kappa_1-\kappa_1^0$ is plotted and 
all other parameters are chosen as above).  
While the main dynamics is captured by the model, neglecting the spatial correlations shows 
here in smaller accuracy after a few tens of iterations of the model, as compared to 
the previous cases. 
This points at the limits of the present correlation model 
when compared with the model in \cite{pbs2} explicitly calculating fluctuations from spin correlations. 
However, the simplicity of the correlation model makes it applicable to 
fitness landscapes where maximum entropy calculations are not feasible. 
Finally, the genetic algorithm run demonstrates that recombination
is no guarantee for improved optimization as long as the encoding does not reward 
with improved correlation. 

\section{Conclusions}  
A dynamical model for the mutation and recombination operators of genetic algorithms 
has been developed, based on a simple correlation measure. The motivation was the
common intuition that the fitness correlation between parents
and children is a measure for the convergence properties of genetic algorithms.  
The correlation determines the model for a microscopic mutation and recombination event, 
which is used as an input for a dynamical formalism of genetic algorithms.  
For two test functions, an additive, random field paramagnet  
and the spin glass motivated NK-model, the dynamics of a genetic algorithm has been modeled and 
compared to the average dynamics of an ensemble of real genetic algorithm runs.  

Three main results can be summarized from this study. 
First, the correlation model helped us in understanding the improved genetic algorithm 
performance on correlated landscapes. 
Second, we obtained a simple model for genetic algorithm dynamics 
on the basis of fitness correlations. A comparison to 
a more involved maximum entropy model \cite{pbs2}  
demonstrated that for the cases considered main features of the dynamics are already contained 
in the fitness correlations of the genetic operators. This gives a simple model at hand 
for fitness landscapes where maximum entropy calculations are not feasible 
as in many practical applications. 
Third, we demonstrated a working model of genetic algorithm dynamics 
on a hard optimization problem, the rugged fitness landscape of the NK model. 

A further goal of this study was to link fitness correlation measures, which are often 
used as empirical measures for genetic algorithm performance,
to dynamical models of genetic algorithms. 
This touches the issue of choosing the right algorithm for a given problem and the 
question which probes might help in this decision \cite{nfl}. 
Fitness correlation measures are among the candidates for such probes. 

\acknowledgments
The author thanks the Deutsche Forschungsgemeinschaft for funding this study.

\begin{center}  
{\bf FIGURES}  
\medskip
\end{center}
 
\noindent   
\begin{figure}[h]
\caption{Measured evolution (solid lines) and predicted evolution (dashed lines) 
of $\kappa_1$ and $\kappa_2$ for a random field paramagnet 
fitness under selection and mutation.}   
\label{rp1}
\end{figure}

\noindent   
\begin{figure}[h]
\caption{Measured and predicted evolution of 
$\kappa_1 - \kappa_1^0$ and $\kappa_2$ for the NK-model fitness 
under selection and mutation.}   
\label{nk1}
\end{figure}

\noindent   
\begin{figure}[h]
\caption{Measured and predicted evolution of $\kappa_1$ and $\kappa_2$ for the random field paramagnet 
fitness under selection, mutation, and recombination.}   
\label{rp2}
\end{figure}

\noindent   
\begin{figure}[h]
\caption{Measured and predicted evolution of $\kappa_1 - \kappa_1^0$ and $\kappa_2$ 
for the NK-model fitness under selection, mutation, and recombination. 
}   
\label{nk2}
\end{figure}

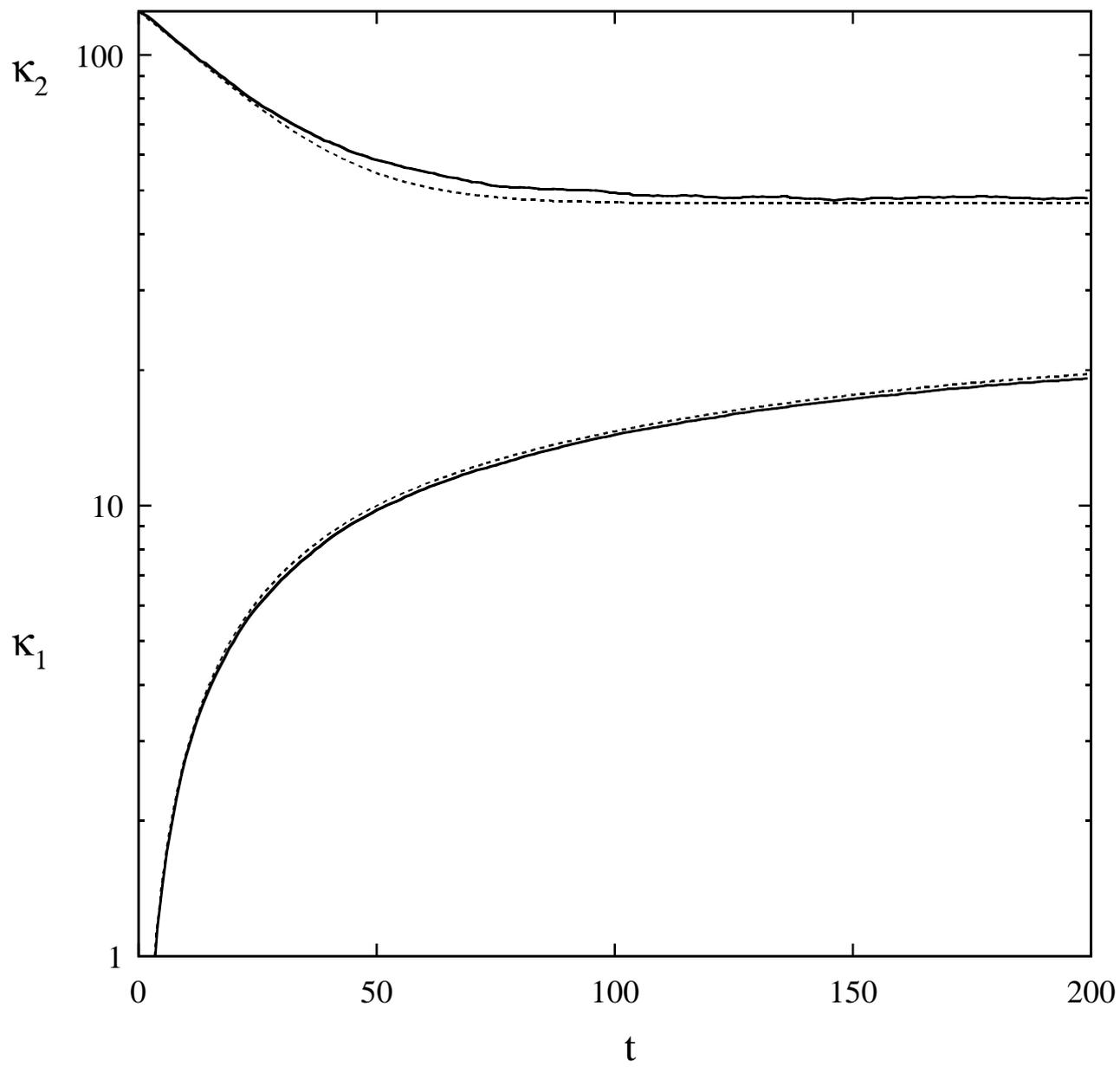
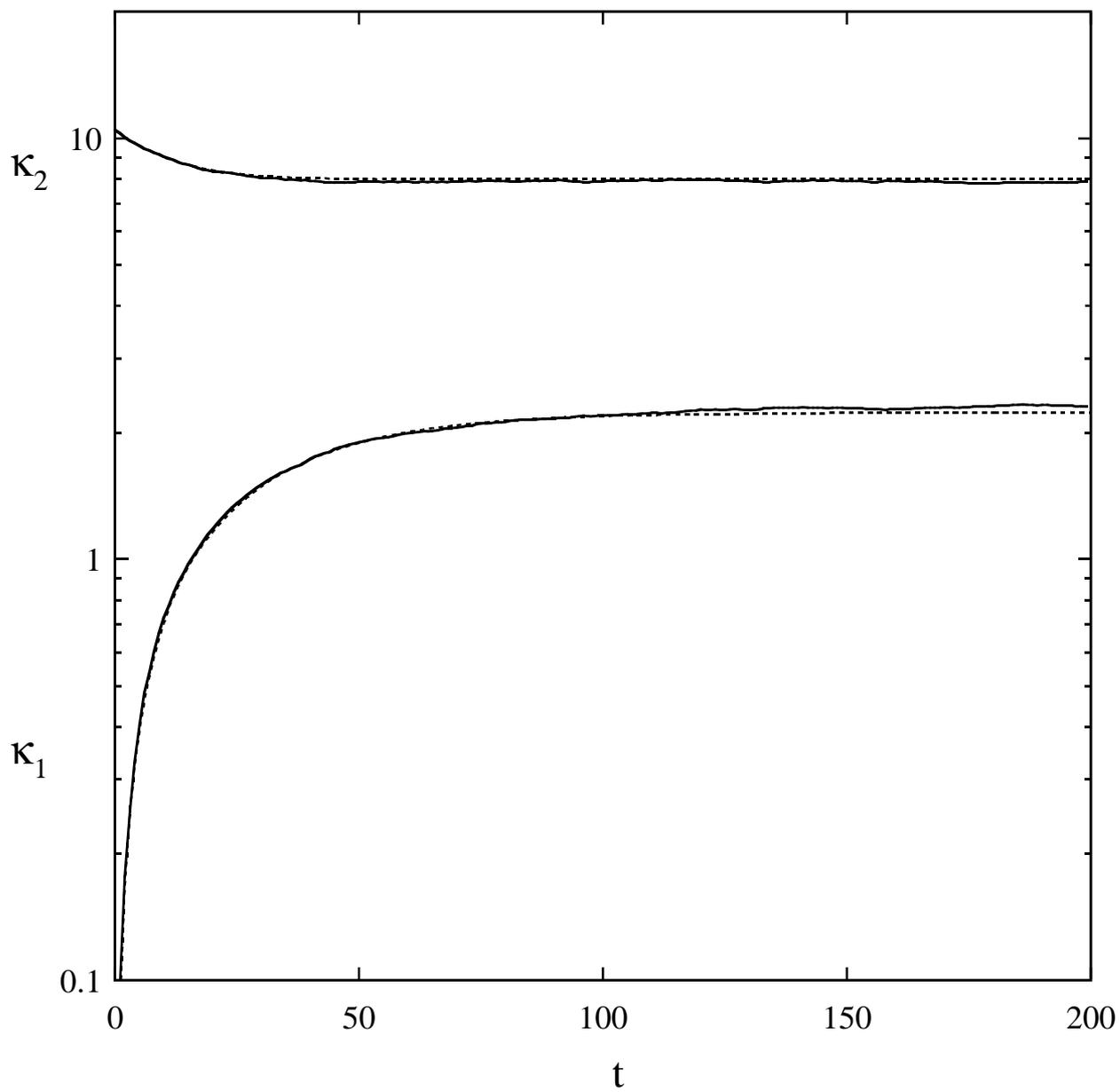
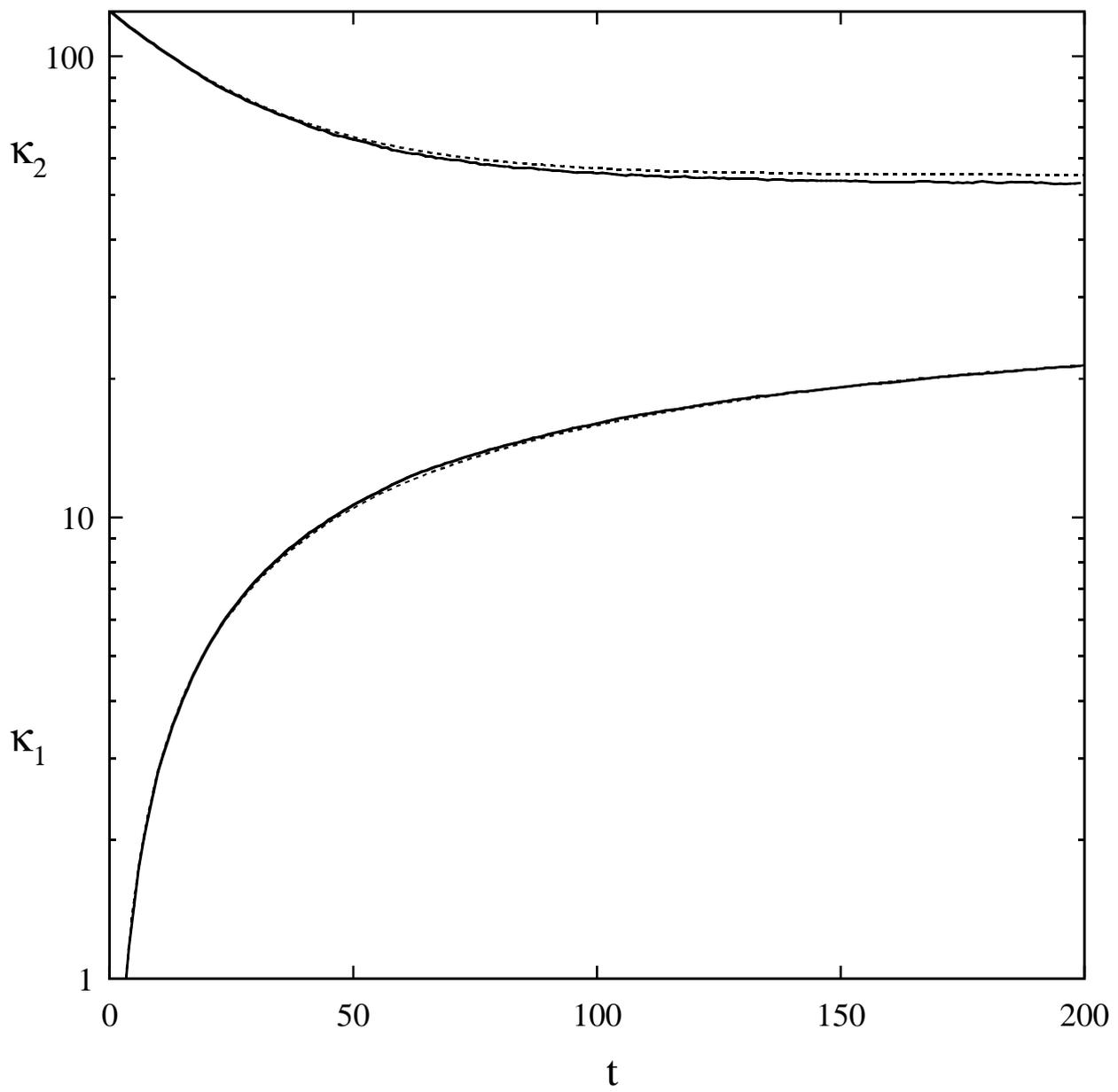
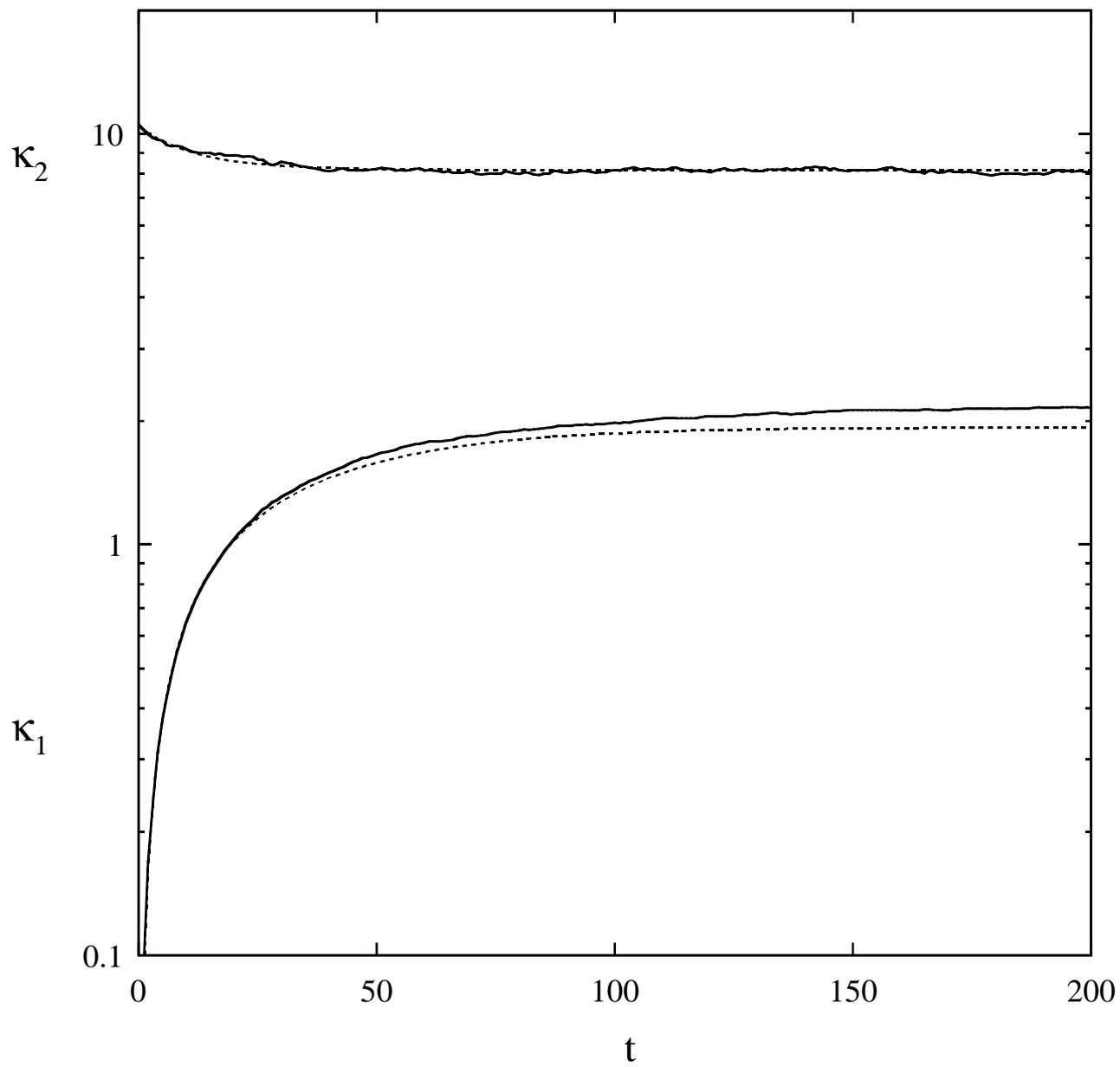
\end{document}